\documentclass[twocolumn,aps,pra,superscriptaddress,amsmath,amssym,floatfix,longbibliography,notitlepage,amsbook]{revtex4}
\usepackage[colorlinks=true,linkcolor=blue,citecolor=blue,filecolor=green,urlcolor=blue]{hyperref}
\usepackage{graphicx}
\usepackage{algorithm}
\usepackage{algpseudocode}
\usepackage{physics}

\def\be{\begin{equation}}
\def\ee{\end{equation}}

\begin{document}

\newcommand{\indep}{\perp \!\!\! \perp}

%Title of paper
\title{Raspberry Pi multispectral imaging camera system (PiMICS): a low-cost, skills-based physics educational tool.}

\author{John C. Howell}
\affiliation{Institute for Quantum Studies, Chapman University, Orange, CA 92866, USA}
\affiliation{Racah Institute of Physics, The Hebrew University of Jerusalem, Jerusalem, Israel, 91904}

\author{Brian Flores}
\affiliation{Institute for Quantum Studies, Chapman University, Orange, CA 92866, USA}
\author{Juan Javier Naranjo}
\affiliation{Department of Physics, Escuela Politécnica Nacional, Quito, Pichincha 170525, Ecuador.}

\author{Angel Mendez}
\affiliation{Department of Physics, Escuela Politécnica Nacional, Quito, Pichincha 170525, Ecuador.}

\author{César Costa-Vera}
\affiliation{Department of Physics, Escuela Politécnica Nacional, Quito, Pichincha 170525, Ecuador.}

\author{Chris Koumriqian}
\affiliation{Institute for Quantum Studies, Chapman University, Orange, CA 92866, USA}

\author{Juliana Jordan}
\affiliation{Institute for Quantum Studies, Chapman University, Orange, CA 92866, USA}

\author{Pieter H. Neethling}
\affiliation{Stellenbosch Photonics Institute, Stellenbosch University, Stellenbosch South Africa}
\affiliation{National Institute for Theoretical and Computational Sciences, South Africa}

\author{Calvin Groenewald}
\affiliation{Stellenbosch Photonics Institute, Stellenbosch University, Stellenbosch South Africa}

\author{Michael A. C. Lovemore}
\affiliation{Department of Physics, University of Pretoria, Pretoria, South Africa}

\author{Patrick A. T. Kinsey}
\affiliation{Department of Physics, University of Pretoria, Pretoria, South Africa}

\author{Tjaart P. J. Krüger}
\affiliation{Department of Physics, University of Pretoria, Pretoria, South Africa}
\affiliation{National Institute for Theoretical and Computational Sciences, South Africa}

\email{johhowell@chapman.edu}

\begin{abstract}
We report on an educational pilot program for low-cost physics experimentation run in Ecuador, South Africa, and the United States.  The program was developed after having needs-based discussions with African educators, researchers, and leaders.  It was determined that the need and desire for low-cost, skills-building, and active-learning tools is very high.  From this, we developed a 3D-printable, Raspberry Pi-based multispectral camera (15 to 25 spectral channels in the visible and near-IR) for as little as \$100.  The program allows students to learn 3D modeling, 3D printing, feedback, control, image analysis, Python programming, systems integration and artificial intelligence as well as spectroscopy.  After completing their cameras, the students in the program studied plant health, plant stress, post-harvest fruit ripeness, and polarization and spectral analysis of nanostructured insect wings, the latter of which won the ``best-applied research" award at a conference poster session and will be highlighted in this paper.  Importantly, these cameras can be an integral part of any developing country's agricultural, recycling, medical, and pharmaceutical infrastructure. Thus, we believe this experiment can play an important role at the intersection of student training and developing countries' capacity building.     
\end{abstract}

\maketitle
Multispectral (MSI) and hyperspectral imaging (HSI) have proven to be invaluable resources in precision agriculture, environmental monitoring, mineralogy, geology, food quality and inspection, forensics, chemical analysis, medicine, recycling, defense, and planetary sciences \cite{van2012multi,zhang2012application,wu2013advanced,qin2013hyperspectral,mansfield2014multispectral,fei2019hyperspectral,nalepa2021recent,adao2017hyperspectral,calvini2019growing,ortega2019use,lu2020recent,peyghambari2021hyperspectral,de2023hyperspectral}. While MSI captures information from discrete spectral bands, providing targeted spectral information useful for cost-effective applications, HSI captures hundreds of contiguous bands, offering detailed spectral data across a broader range \cite{adao2017hyperspectral,nalepa2021recent}. Both HSI and MSI techniques have been instrumental in agricultural advancements \cite{adao2017hyperspectral,lu2020recent}, where they facilitate crop health monitoring, disease detection, and soil analysis for improved water, pesticide, and fertilizer usage. In healthcare, MSI and HSI have been adopted for non-invasive diagnostics \cite{mansfield2014multispectral,fei2019hyperspectral,ortega2019use}, particularly in cancer detection and surgical guidance, where they exploit subtle spectral variations between healthy and pathological tissues. Environmental monitoring applications \cite{zhang2012application} leverage these technologies for assessing land cover, water quality, and pollutant detection, aiding in sustainable ecosystem management. Industrial applications, such as quality control and food safety \cite{qin2013hyperspectral}, also benefit from these imaging modalities, where they allow for rapid, non-destructive assessments based on chemical composition and structural integrity. Recent advancements in sensor technology and machine learning for MSI and HSI data processing are expanding the utility of these techniques in real-time, high-accuracy applications across diverse sectors.  

It should be clear that these cameras can or do play an important role in bringing added value and capabilities to a society.  However, one of the problems with the uptake of these cameras is that they can be cost-prohibitive, especially in developing countries. The purpose of this paper is to share a pilot program that trains students in developing countries to build low cost, size, weight and power (C-SWaP) multi-spectral cameras, enabling them to further innovate and develop this powerful tool.

\section{Background}
John Howell, the lead author of this paper, was elected president of the International Commission for Optics (ICO) in 2021 for a three-year term.  After starting his term, Howell formed an African Advisory Committee to the ICO composed of physics, optics and photonics researchers and educators from several countries across Africa.  Several needs were discussed and perspectives were shared.  Of the elements discussed, hands-on active-learning experimentation, computer skills and physics/optics capacity building in the form of student capabilities were targets of interest for the ICO mission.

After the meetings, it was determined that a low-cost multispectral imaging camera system could play an important role at the intersection of education and critical infrastructure as discussed earlier. A Raspberry Pi-based multispectral imaging camera system (PiMICS) was developed.  The purpose of the program is two-fold: 1) provide a framework for gaining many important physics and engineering skills, and 2) train the students in an important technological domain which can have significant economic and quality-of-life implications for their respective countries.  The PiMICS program can further aid countries in meeting a host of sustainable development goals: increasing crop yields, reducing water consumption, monitoring water pollution and purity, and improving medicine, to name a few. This program solves the serious problem in developing countries where the cost of MSI and HSI cameras is simply too high for the average researcher. The architecture, a Raspberry Pi 4b, was chosen for its cost, size, weight and power allowing it to be portable, but still computationally capable.  The Raspberry Pi can act as the control mechanism, the processor, and the camera.  

\section{Raspberry Pi-Based System}
The Raspberry Pi 4b has many features that make it ideal as an instructional tool.  As a note, the authors believe that the Pi 5 consumes too much power to make it a viable portable device as most off-the-shelf 5-V power banks only supply 3 amps, which can compromise peripherals (necessary for this project).  The Pi 4b is a quad-core, 1.5-GHz single-board computer.  It has up to 8 GB of SDRAM, Bluetooth, 2 USB2 ports, 2 USB3 ports, a gigabit Ethernet port, wifi, and allows remote access through programs like RealVNC or TeamViewer. The Pi 4b can run up to two 4k monitors. It has a dedicated camera serial interface port connected through a 15-pin cable. The Camera Module 3 NoIR can take HD 1080p video or 12-megapixel still images in the visible and near-infrared.   Importantly, the Pi 4 has a 40-pin GPIO header with 26 GPIO pins. The GPIO pins can control motors and LEDs.  It is important to note that one should not power the motors or LEDs through the Pi header, but through a separate supply to prevent damaging the GPIO header.  Combined with Python software pre-installed on the Pi 4 (many of the packages used in Python are not pre-installed), it provides a powerful interface between programming and experimental control and feedback.  Adding to the portability of the system is the fact that there are inexpensive 4-inch (13 cm), 5-inch (15 cm), and 7-inch (18 cm) monitors that can be mounted directly to the Pi mitigating the need for remote or wired monitor connections while being operated.  We used the 4-inch monitor for PiMICS Zero.  It is adequate to run or operate programs, but too small for programming.  The 5-inch display is probably better, but we are concerned that the 7-inch display might draw too much power (potentially up to 6 W) considering the heavy load of all the peripherals during the experiment.  

\subsection{3D printing}
The compact size of the Raspberry Pi also makes it ideal for use with 3D printing.  Lightweight, inexpensive, purpose-built, and/or aesthetic 3D architectures can be designed and printed according to experimental needs.  The architectures can incorporate onboard power, light-emitting diodes, motor control and so forth. Free 3D modeling packages include Fusion360, Blender, Tinkercad, etc.  For this project, Fusion360 and Blender were used to create stereo lithographic (STL) files and which were then imported and sliced in PrusaSlicer generating gcode and bgcode files.  Prusa MK4S and Prusa Mini 3D printers were used to print the camera bodies.  At the time of authoring this paper, a partially-assembled Prusa mini costs less than \$500.  This is a very reliable and capable 3D printer.  The STL printer files for this project are available on GitHub. 

\subsection{Pi 4 Python Programming}
Thonny, the pre-installed integrated development environment (IDE) for Python, allows the students to write and execute Python code on the Raspberry Pi.  We note that other IDEs can be used, but the user must be aware that they may operate relatively slowly on the Pi.  We used both Thonny and VS code.  We found that Thonny worked in all instances, but lacks some of the bells and whistles of other IDEs.  From the Python program, it is possible to drive LEDs, measure reflectance spectra in the camera, perform image processing and analysis of the image, and provide feedback to the system.  The students utilize several Python packages NumPy (numerical processing), Matplotlib (plotting and graphing), time (temporal control of the program), OpenCV (computer vision package with power image processing and machine learning tools), Adafruit ServoKit (motion control), GPIO Zero (LED control) as basics and many more depending on needs. Several levels of complexity of code are available to be used are also in Github.  

\subsection{Active Illumination: Light Emitting Diode System}
In the most basic multispectral camera, the GPIO pins drive individual LEDs of different wavelengths.  LEDs provide for low-cost narrowband illumination ideal for many multispectral imaging applications and have been in common use for multispectral imaging as they have continued to develop over time \cite{li2012multi,shrestha2013multispectral,salvador2018low}.  With the myriad of colors available, a relatively fine-scale spectral probe of the reflectance of an object can be made.  Two different versions of LEDs were used during the program.  In the simplest configuration, standard 5-mm LED packages (19 different colors) with a maximum current of approximately 20 mA were used. DigiKey is a good resource for finding 5-mm LEDs spanning the range of the visible and near-infrared.  Likely owing to the smaller bandgap, it was easier to 5-mm packages of many different colors in the infrared than in the visible.  Of course, many vendors will sell seven-color 5-mm LED packages in the visible wavelengths, but it was difficult to fill in the spectral gaps beyond the seven colors. The 5-mm packaged LEDs have built-in lenses in an effort to concentrate the light in the forward direction.  However, the illumination patterns can vary drastically between LED colors and they tend to have intensity rings, making them more difficult to calibrate spatially. 

The second type of LEDs used in the program were 3-W chip-on-board (COB) LEDs (14 or 15 colors were used).  They are easily found on Amazon.com, Inc.  They were run well below 1 W for safety and power reasons.  For precision spectroscopy, the 3-W LEDs are greatly preferred, because their illumination is much more uniform and they have significantly increased light flux. Uniform illumination makes it much easier to calibrate reflectance flux relative to a Teflon reference target.  

\subsubsection{LED driving circuit}
\begin{figure}[ht]
    \includegraphics[width=.40\textwidth]{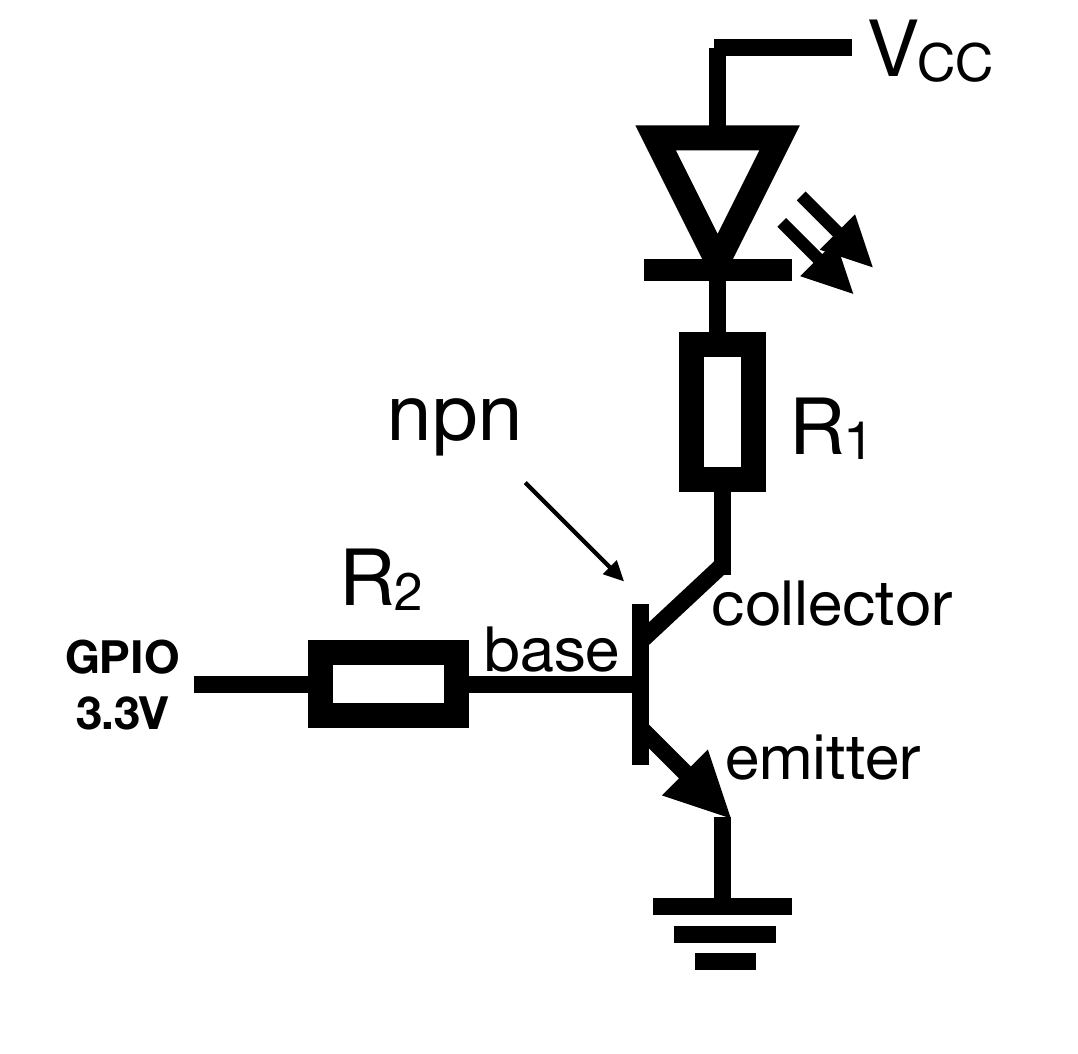}
    \caption{Experimental schematic of the basic circuit to drive a single LED}
    \label{circuit}    
\end{figure}

A simple circuit to drive the LEDs is shown in fig. \ref{circuit}. An npn transistor, driven by a GPIO pin, controls the state of the LED operated in common emitter mode.  As an important note, it is trivial to directly connect low-current LEDs in series with an appropriate resistor to drive the LEDs directly from the GPIO pins rather than use a transistor setup as shown in the figure.  However, it is discouraged to do this for a few reasons.  First, students should learn the power and versatility of using switching transistors.  Transistors have brought us modern information and computational revolution after all.  Second, it protects the GPIO board and reduces the power drawn from the Pi.   The maximum current that should be continuously drawn from a single GPIO pin is about 16 mA and a maximum of about 50 mA from all pins.  If one accidentally turns on all pins or puts in the wrong resistor, it can irreparably damage the GPIO header!  So, even though it is slightly more expensive and is a little more involved to use an additional voltage source like a 9-V battery along with additional transistors and resistors, in the long run, it will likely be cheaper.  Third, it is easy to upgrade the LEDs to higher-power versions as was done in the upgraded version. 

The resistors $R_1$ and $R_2$ shown schematically in fig. \ref{circuit} should be chosen somewhat carefully in concert with the transistor.  A generic 2n2222 npn transistor (peak collector current is about 800 mA) was used, although any suitable npn transistor will work. For a few dollars, it is possible to buy hundreds of these transistors. To determine the resistors used, start by considering the desired collector current $I_C$.  The collector current is amplified by the base current given by $I_C=\beta I_B$.  The value of $\beta$ depends on the collector current, but for collector currents above 10 mA, we will assume $\beta=100$.  This is not too critical because the circuit will often be operated in saturation mode, where increasing current on the base does not increase the collector current.  Assuming that $\beta=100$ and desiring 20 mA in the LED, we then see that we need the base current to exceed 0.2 mA coming from the GPIO pin.  Assuming that the transistor is saturated, implying that the base emitter voltage is $V_{BE}=0.7$ V, it is found that
\begin{equation}
    R_2=\frac{V_{GPIO}-V_{BE}}{I_B},
\end{equation}
which equals 13 k$\Omega$.  However, there is a 1/10 rule-of-thumb in transistor saturation.  Essentially, excess current on the base drives the transistor to saturation.  Increasing the base current beyond saturation does not increase the collector current.  In other words, if the base is driven at 1/10 of the maximum collector current, it is essentially guaranteed to be saturated.   So, in this case, anything around 1 k$\Omega$ resistor $R_2$ can be used.  

Now consider $R_1$.  In saturation, the collector-emitter voltage drop $V_{CE}$ ranges between 0.2 V and 0.4 V.  The voltage drop across an LED $V_{LED}$ can range between 1.8 V and 3.4 V. Assuming that the transistor is saturated, it is found that
\begin{equation}
    R_1=\frac{V_{CC}-V_{LED}-V_{CE}}{I_C}.
\end{equation}
Using a battery with $V_{CC}=9$ V and assuming $V_{LED}=3.4$ V, $V_{CE}=0.2$ V, and $I_C=20$ mA, it is found that $R_1=270\:\Omega$.

In this low-power LED scenario, the total power dissipated power through resistive heating in $R_1$ is $(0.02\textnormal{ A})^2\: 270\: \Omega\approx 0.1$ W.  Thus, 1/4-W resistors are sufficient.  However, if higher-power LEDs are used, it is important to use higher-power resistors. 

For the 3-W LED system, a 6-V battery (4 rechargeable AA batteries in series) was used along with $R_1=22\:\Omega$ and $R_2=1\textnormal{ k}\Omega$ (resistors between 220 $\Omega$ and 1 k$\Omega$ were tried and they all worked well).  The 3-W LEDs require soldering wires to the leads of the LED to fit into a breadboard.  All LEDs were tied to $V_{CC}$ and each base had its unique GPIO pin on the Pi.  A common mistake is to forget to tie the ground of the $V_{CC}$ battery to the ground of the GPIO, which should be done on the breadboard or printed circuit board.  For the high-power system, all components were soldered to a printed circuit board.  

\subsection{Passive Illumination: Narrowband Interference Filters}
\subsubsection{Motor Control}
The passive spectral measurements are performed using a rotating wheel of narrowband interference filters.  The filter wheel was printed and then rotated by motorized control.  During the pilot, both stepper motors (28BYJ48) and servomotors (plastic gears SG90 and metal gears MG90) were used for rotation.  Both the stepper motors and servomotors have advantages and disadvantages.  The stepper motor is easier to use and requires less programming to enable.  It can perform continuous rotations.  The disadvantage is that it requires 4 GPIO pins to run, is slow, and does not have a stable reference point often requiring recalibration.  For the servomotors (MG90 preferred), it was found that the servomotors can be unstable unless an I2C-enabled driver was used.  The HiLetgo PCA9685 driver along with the Adafruit ServoKit Python package were used to drive the servomotors.  This configuration only requires 2 GPIO pins to run up to 16 servomotors with 12 bits of precision!  Without this system, the servomotors can be somewhat unstable and finicky.  As a note, the Adafruit ServoKit Python package required making a separate virtual environment within Python that needed to borrow global dependencies of higher-level programs like PiCamera.  ChatGPT can guide students through the process of solving this issue.  The driver sends 50-Hz (20-ms period) pulse wave modulation (PWM) signals based on the duty cycle (max range 0.6 ms to 2.4 ms). It can take some adjustments to the duty cycle to get the servomotors to behave according to specifications but exceeding the maximum must be avoided to prevent damage.  Thus, the servomotor option can be tricky for students who are just learning Python programming.  One last drawback is that a typical SG90 or MG90 servomotors can only operate through 180 degrees.  However, in the long run, it is recommended to use the servomotors for reasons of speed, repeatability and the ability to expand.  In either case, the motors should be run by an external 5-V power supply and NOT from a 5-V power GPIO pin on the Raspberry Pi, in spite of what might be demonstrated on YouTube videos!    

\subsubsection{Chroma Filters}
\begin{figure}[ht]
    \includegraphics[width=.49\textwidth]{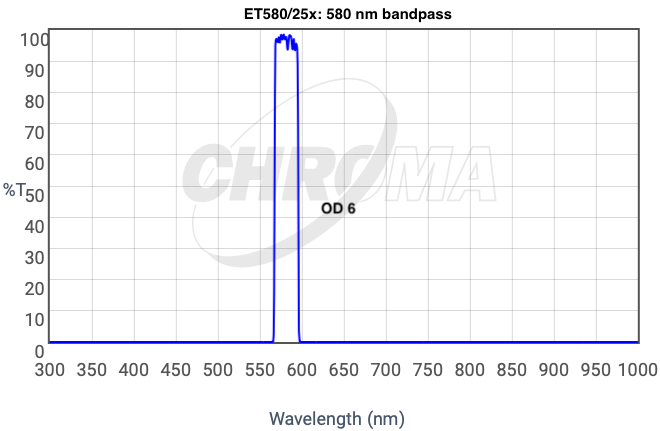}
    \caption{Typical transmission spectrum for the interference filters purchased from Chroma Technology.}
    \label{580Filter}    
\end{figure}

We used 15 Chroma Technology filters, most of which have at least 5 OD extinction outside the passband over the entire measured spectral response of the camera.  A typical transmission spectrum is shown in fig. \ref{580Filter}. The downside of the interference filters vs colored glass is the angular dependence of the spectral passband.  However, for a sufficiently narrow field of view, the spectral window does not shift more than a few nm. The circular apertures of the filter wheel also ensure that the maximum off-axis angle of incoming light is about 30 degrees.   

\subsection{Image Analysis}
The imaging system is straightforward: a light source (LEDs or broadband) illuminates an object as well as a Teflon reference target.  The reflected light is measured in the camera without or with a narrowband interference filter if there is an LED or broadband source, respectively. A picture is taken at each wavelength for the LED or with each filter for the broadband source.   The PiCamera2 Python package is used to acquire the image.  The 8-bit Red Green Blue (RGB) values are then determined for each pixel.  The average 8-bit RGB value of many pixels of the Teflon target is calculated.  The ratio of the pixel values on the object to the averaged pixel values of the Teflon target is then the calculated reflectance for the wavelength probed. As a note, if a student wants to use OpenCV (the Python computer vision package), he/she will first need to convert the image from RGB to BGR. 

The reflectance measurement, described above, is over-simplified.  The underlying assumption is that there is uniform illumination.  If the flux density hitting the object is different from the  Teflon target, the ratio is inaccurate.  This can become important if the location for each LED is different.  This issue can be resolved by moving the LEDs when they are lit so that they only illuminate from one position.  For stationary LEDs, one can limit the non-uniformity effects using chip-on-board (COB) LEDs combined with geometry such that the distance from the camera to the object is much greater than the distance between the object and Teflon target.  Lastly, one can simply calibrate the angular flux density across the field by taking a reference image with a large non-glossy (to prevent glare) white paper for each LED.   

%
% Brian
%
\subsection{Camera Module 3 Considerations}
\begin{figure*}
	\centering
	\includegraphics[width=0.3\linewidth]{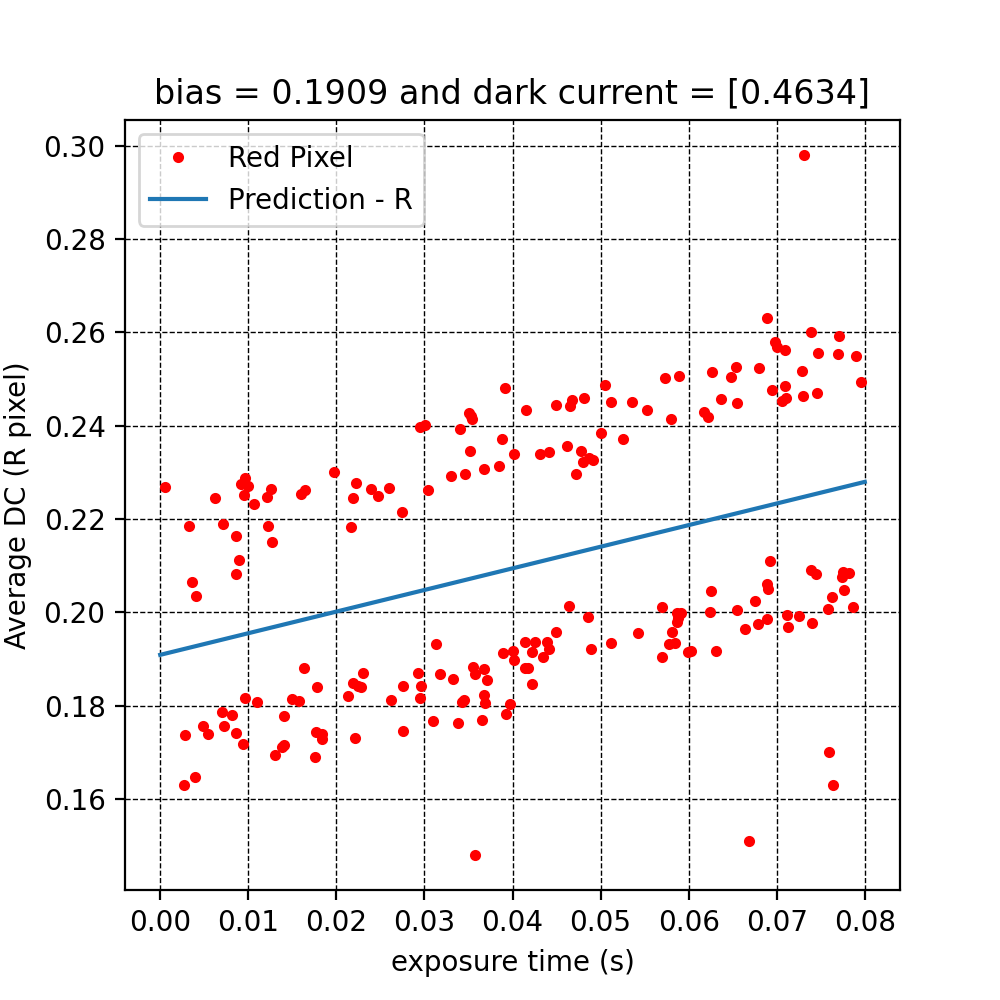}
	\includegraphics[width=0.3\linewidth]{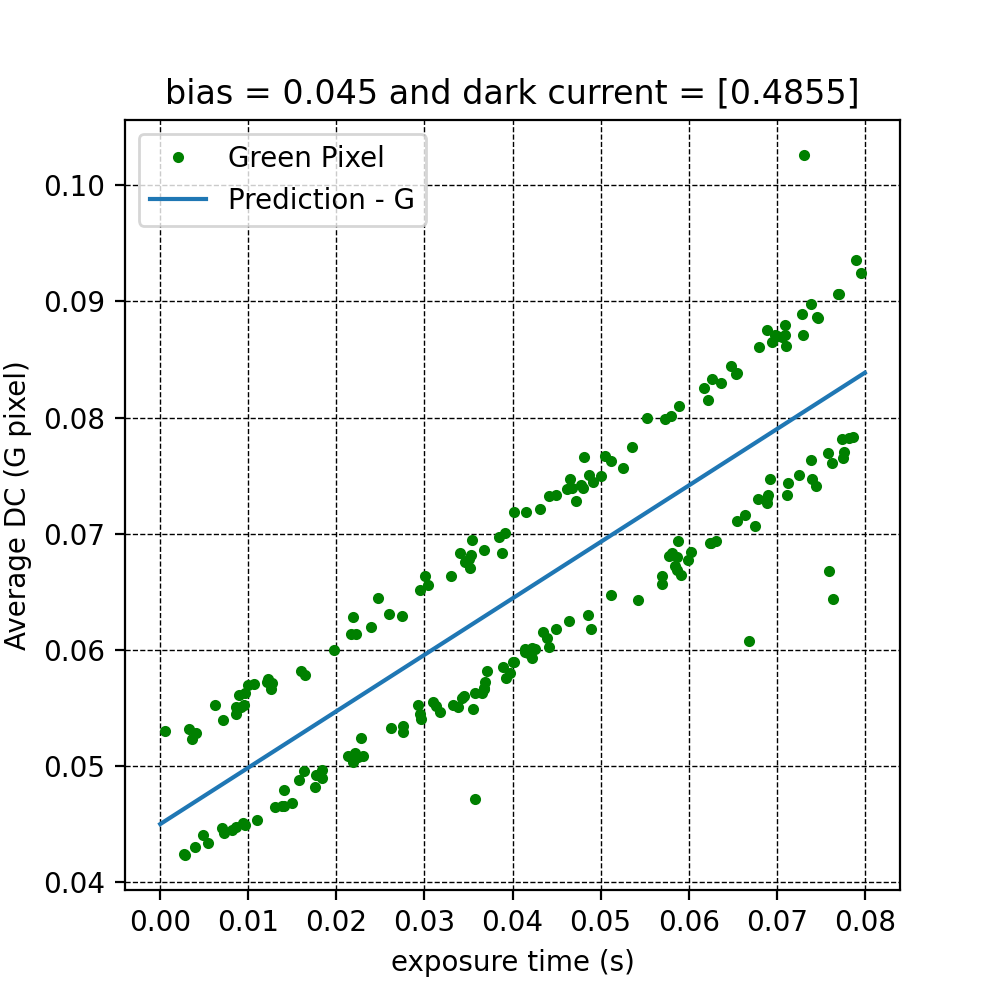}
	\includegraphics[width=0.3\linewidth]{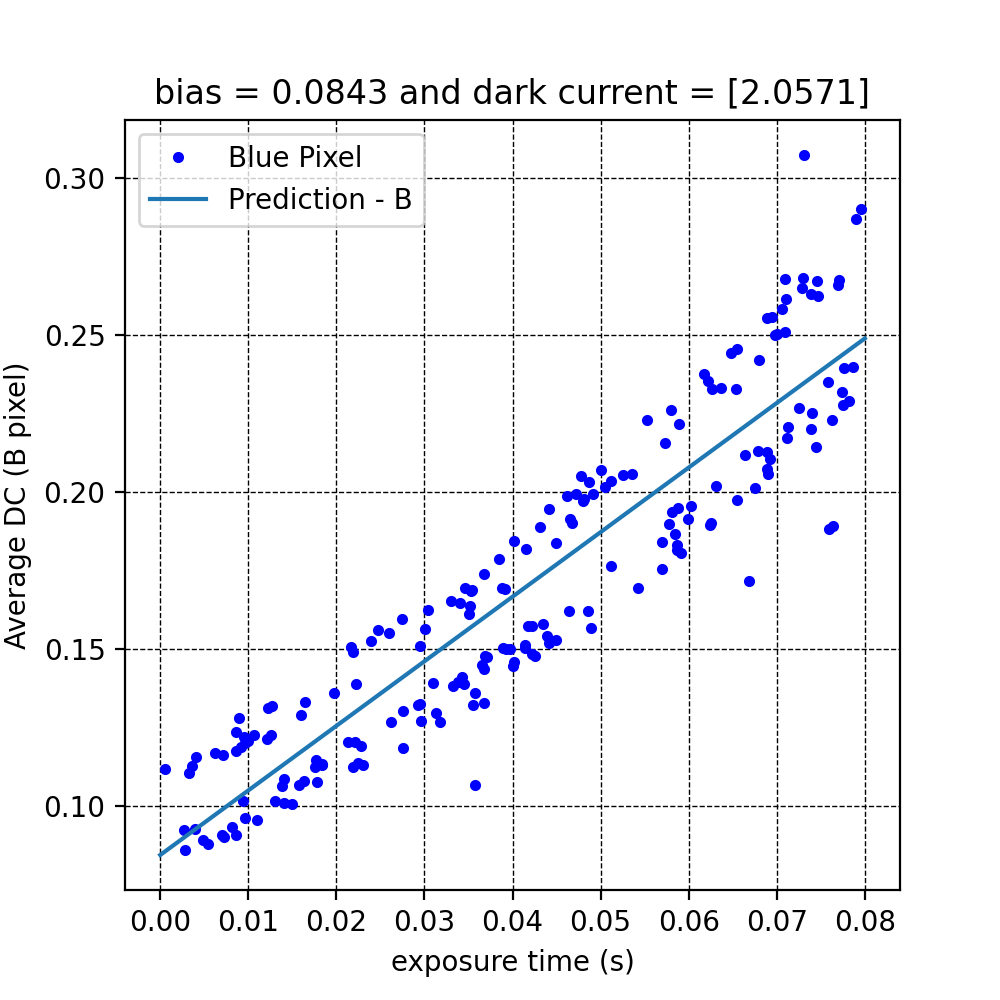}\\
	\includegraphics[width=0.3\linewidth]{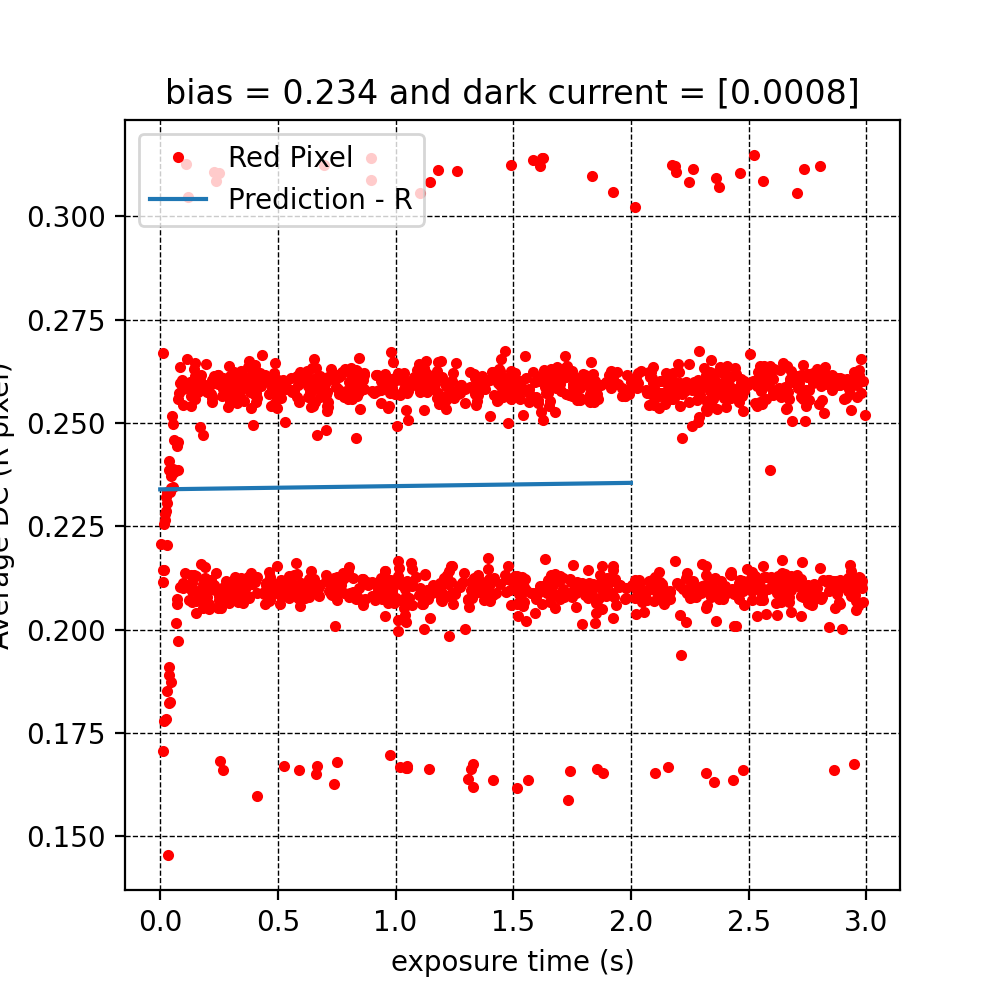}
	\includegraphics[width=0.3\linewidth]{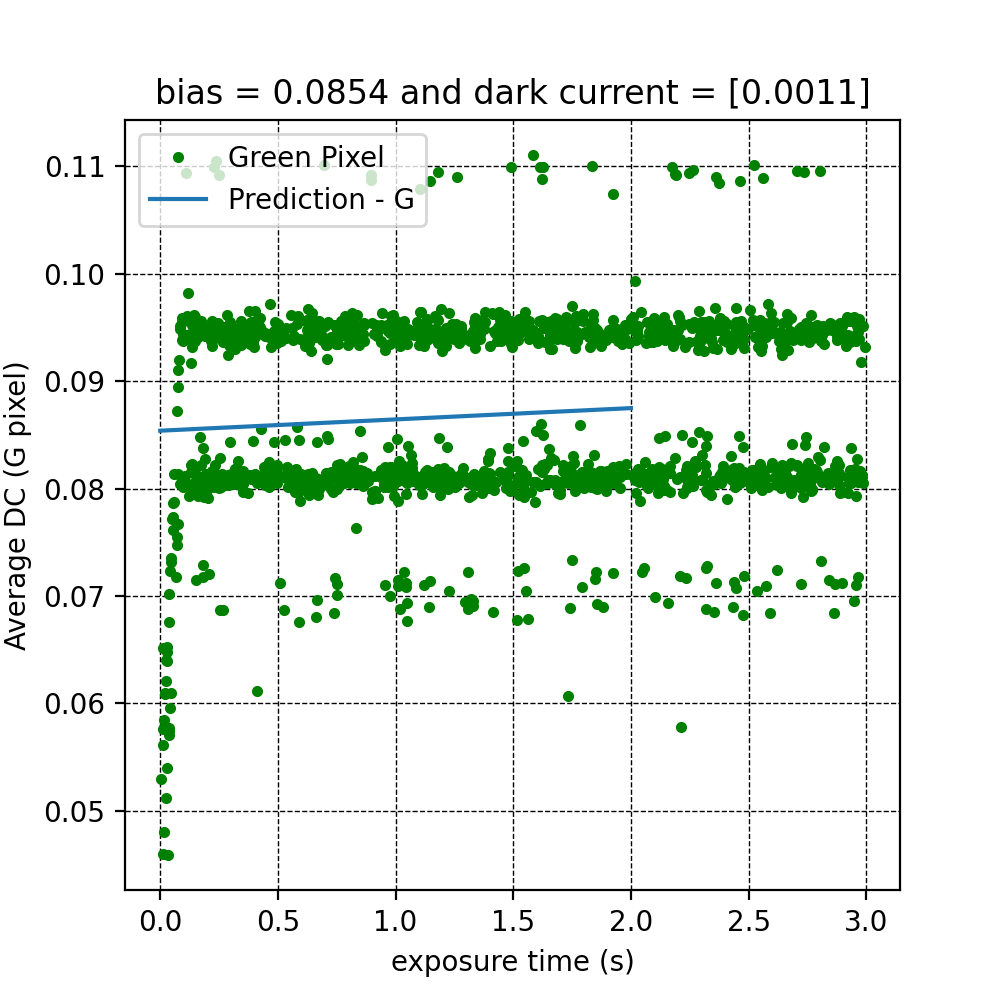}
	\includegraphics[width=0.3\linewidth]{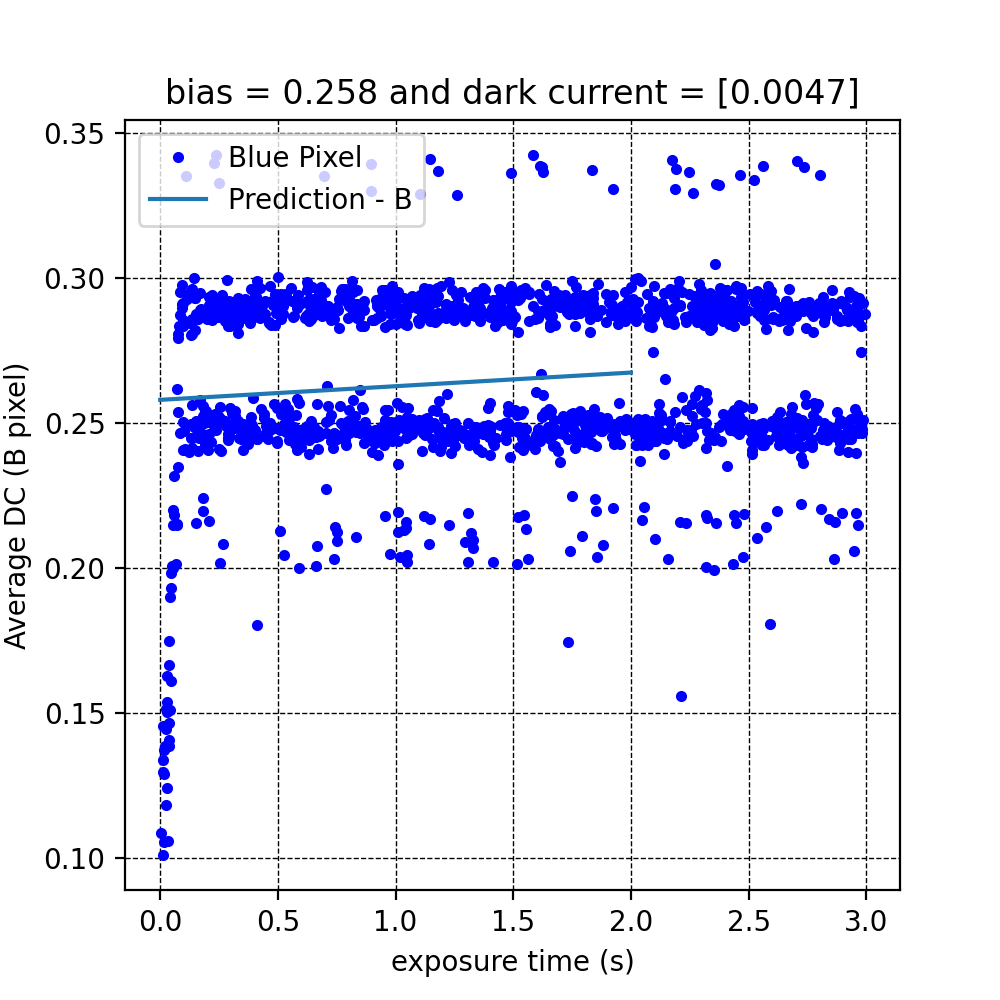}
	\caption{Measured dark current for each RGB pixel, averaged over entire image for various exposure times; linear regression was performed (\textit{solid, blue curve}). The top row shows dark current increasing with longer exposure time; the bottom row shows dark current saturating after $\sim 0.08 s$. }
	\label{Dark_Current}
\end{figure*}

There are important considerations for the Camera Module 3 NoIR.   It is important to set the exposure time of the camera so that the integrated light flux is well below pixel saturation.  This can be done by placing the Teflon target at a given distance from the LEDS and at the center of the illumination, illuminating the LED and changing the camera exposure time until the peak value of any of the RGB pixels from the Teflon is about 200 on a scale of 0 to 255 (8 bits).  It was found that the camera system response can go nonlinear much above this value.   In addition, the red, green, and blue sensors all become sensitive in the near-infrared.  This can create unexpected image colors.  Also, the sensitivity in the near-infrared can be significantly lower than the visible meaning that the exposure time should be calibrated appropriately.  Lastly, we found that the pixels had a small amount of dark current, which grew linearly with exposure time in a short time interval and then saturated.%; see fig. \ref{Dark_Current}. 

\section{Student Engagement and Choice}
Each student or student group chose the experiment they wanted to study.  For more information about some of these projects, see the PiMICS.org website \cite{pimics_website}.  They built cameras that were purpose-built for their experiment.  Typically, students opted to work on multispectral imaging topics that were well established in the literature such as plant health, water stress \cite{behmann2014detection}, or fruit ripeness (bananas \cite{rajkumar2012studies}, avocados, apples, and lemons).  The most developed study performed by the students was the polarization and spectral analysis of nanostructured insect wings, which we will highlight later.

\section{PiMICS Zero}
\begin{figure}[ht]
    \includegraphics[width=.47\textwidth]{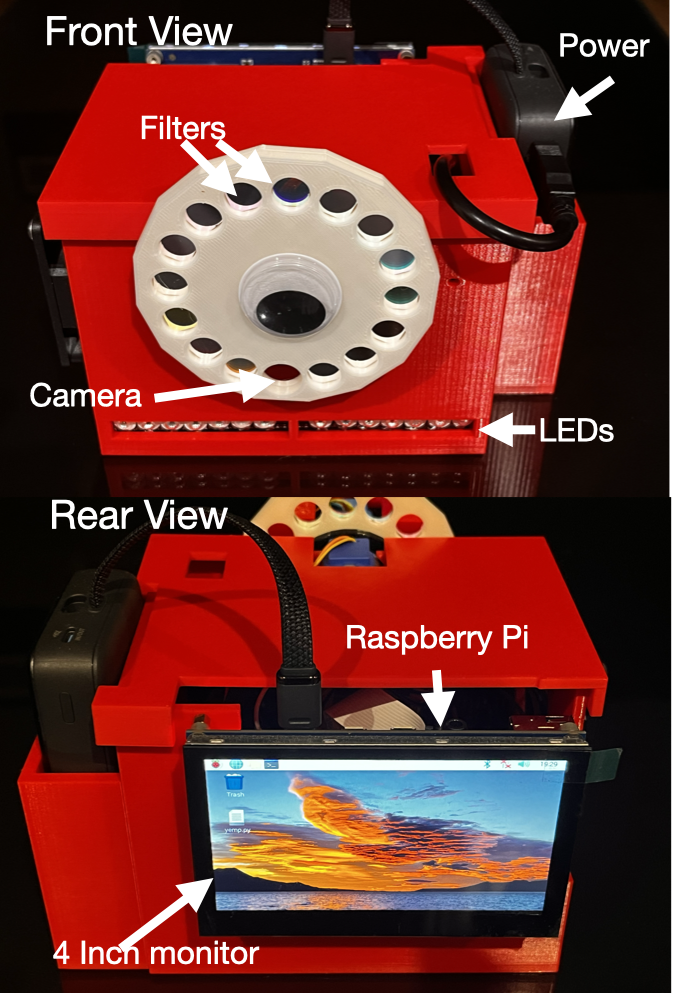}
    \caption{Front and rear views of PiMICS zero.  It has 14 3-W LEDs for active illumination and 15 narrowband filters for passive illumination of objects.}
    \label{Zero}    
\end{figure}

\begin{figure}[ht]
    \includegraphics[width=.49\textwidth]{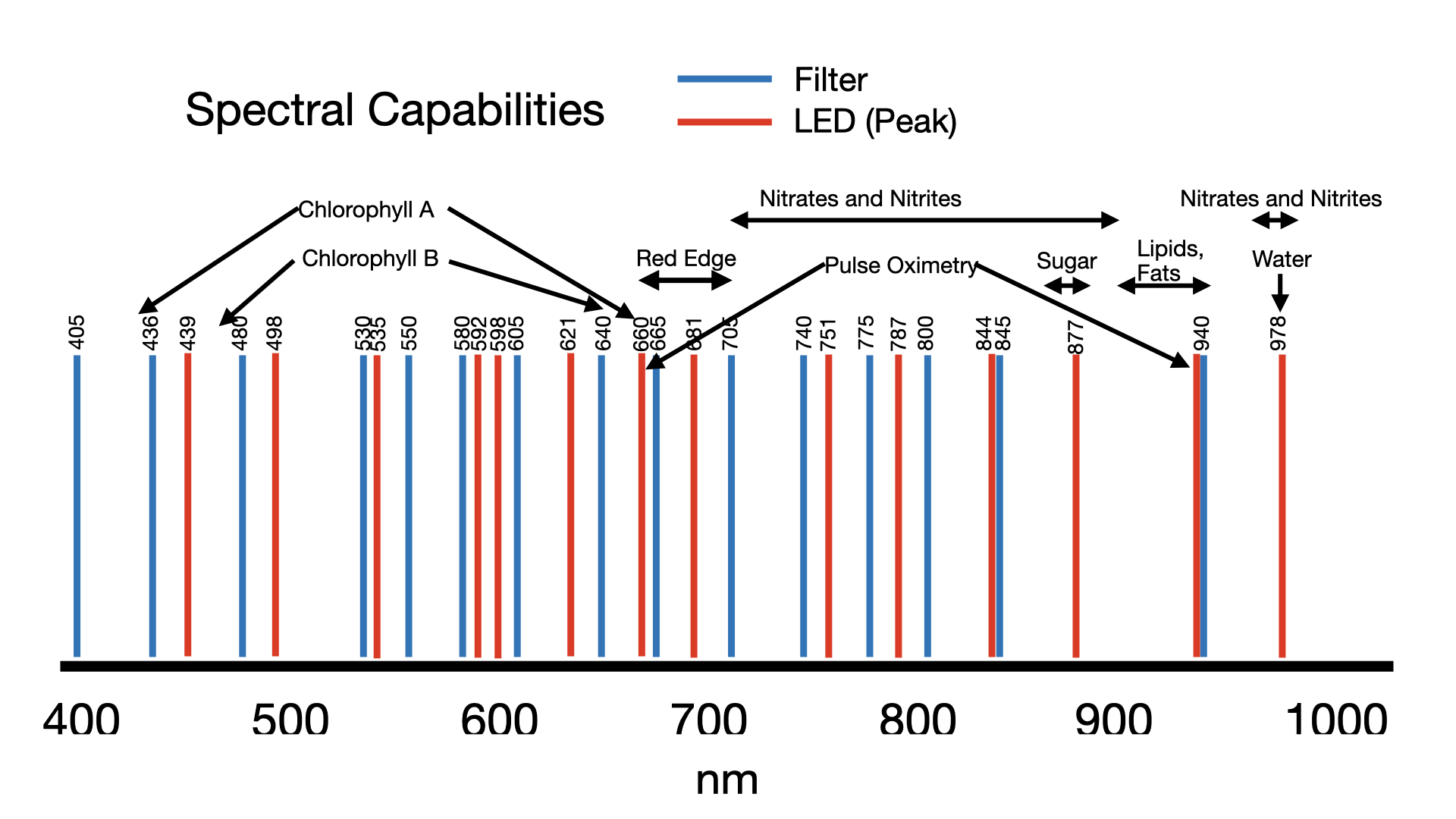}
    \caption{Measured peak wavelength of the LEDs along with the center wavelength of the narrowband filters are shown.  Important resonance spectral features for plants and skin analysis are also shown.   }
    \label{LEDBandSpectra}    
\end{figure}

Some general-purpose and several purpose-built cameras were designed during the pilot program. In particular, the focus will be on two principal cameras.  The first, a general-purpose multispectral camera (both active and passive illumination) for internal historical reasons, is called PiMICS Zero.  It is shown in fig. \ref{Zero}. There are 14 3W COB LEDs, controlled by Raspberry Pi general-purpose input/output (GPIO) pins, shown in the front view towards the bottom. There are also 15 narrowband interference filters placed in a 3D-printed filter wheel along with one additional hole with no filter rotated by the stepper motor described above.  The peak LED wavelengths (as measured in a spectrometer) and center frequencies of the interference filters are shown in fig. \ref{LEDBandSpectra} along with important agricultural and human spectral resonances.  The camera incorporates a 4-inch monitor on the back allowing the user to run programs, process data and view images even while offline.  There is a 30W 10,000mAh power bank that can run the camera for many hours ($\approx$6h depending on use).  The primary power cable (15W)from the power bank runs the Pi and the other cable (15W) runs the motor and the fan.  The use of the fan both cools the Pi and prevents the powerbank port from shutting off when the stepper motor is inactive. 

\section{Multispectral and polarization imaging of butterfly wings}

\begin{figure}[h]
    \centering
    \includegraphics[width=\linewidth]{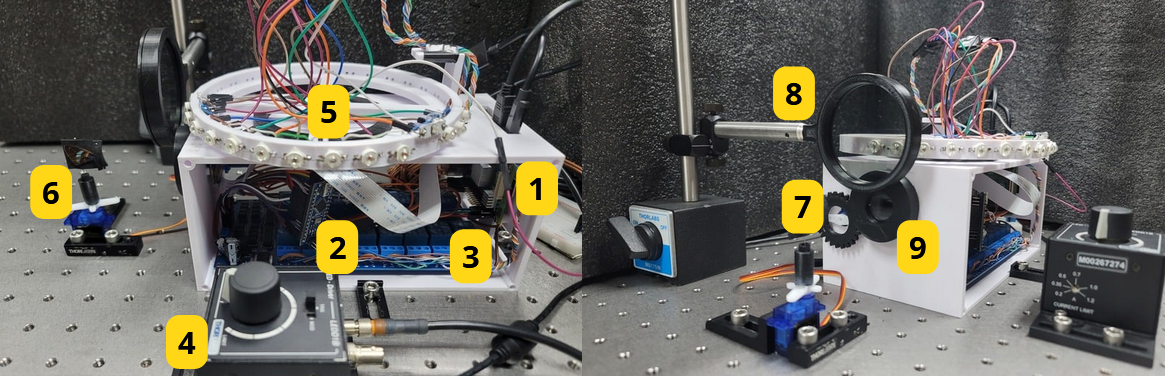}
    \caption{Components of PiMICS for multispectral and polarization imaging. Listed: 1. Raspberry Pi 4 (4GB), 2. Arduino Nano, 3. 16 Relay Module, 4. Thorlabs LED Driver, 5. Motorized LEDs Wheel, 6. Motorized Sample Holder, 7. Motorized Camera Polarizer, 8. Light Source Polarizer, 9. Raspberry Camera Module 3 NoIR.}
    \label{EPN_PiMICS}
\end{figure}

The second camera system presented in this paper was designed to analyze the polarization and spectral properties of nanostructured insect wings. It is shown in fig. \ref{EPN_PiMICS}. 

\begin{figure}
    \centering
    \includegraphics[width=\linewidth]{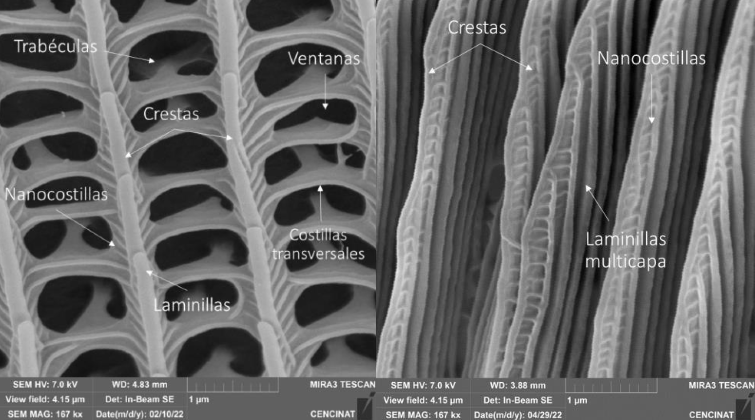}
    \caption{Scanning electron microscopy (SEM) images of butterfly wings microstructures. Non-specialized structures characterized by quasi-parallel ridges along their length, covered by thin lamellae and forming a set of light diffraction planes (left).  Lamellar ridge structures (also called `thin-film reflectors') composed of a series of overlapping lamellae forming multilayer ridges. The reflectance in the ultraviolet (UV) range is enhanced due to constructive interference (right) \cite{LopezAlvarado2022}}.
    \label{wings_sem_EPN}
\end{figure}

The iridescence observed in butterfly wings is the result of light interacting with nanometric structures embedded within the scales, see fig. \ref{wings_sem_EPN}. These structures function as a diffraction grating or interference sheet, by which the thickness of the layers and the configuration of the materials result in the reinforcement or cancelation of specific wavelengths of light, depending on the angle of incidence and observation \cite{wingsiridiscence}. Emergent structural colors are markedly contingent on the geometry and refractive index of the microstructures. In contrast to pigment colors, these colors are not susceptible to fading or discoloration over time, as they depend on shape and not on a chemical reaction.

While there is considerable variation in the nanostructures observed, with or without iridescence, the wings consistently exhibit longitudinal ridges situated parallel to the scale axis and spaced between 500 and 5,000 nm. The optical properties of these wings are highly dependent on the spacing of the ridges and the angle of incidence of light. Furthermore, these configurations could potentially give rise to light polarization, resulting in the observed color contrasts within the species and influencing the reflectance of specific wavelengths \cite{LopezAlvarado2022}. Some species with lamellar ridge-type structures show iridescence with bright blue to green colors, generated by interference reaching high reflectivity \cite{Wickham2005}.

%\section{Multispectral imaging of bananas and lemons}

%In developing countries, it is imperative to quantify how rapidly different food types ripen (this is especially important in hotter climates). For the purpose of this demonstration of the multispectral camera, we investigated the trend between ripeness and average reflectance for lemons and bananas. (still to do - add lemon and banana 3D reflectance spectra and discuss trends )

\subsection{Experimental Setup}

\begin{figure}[h]
    \centering
    \includegraphics[width=1\linewidth]{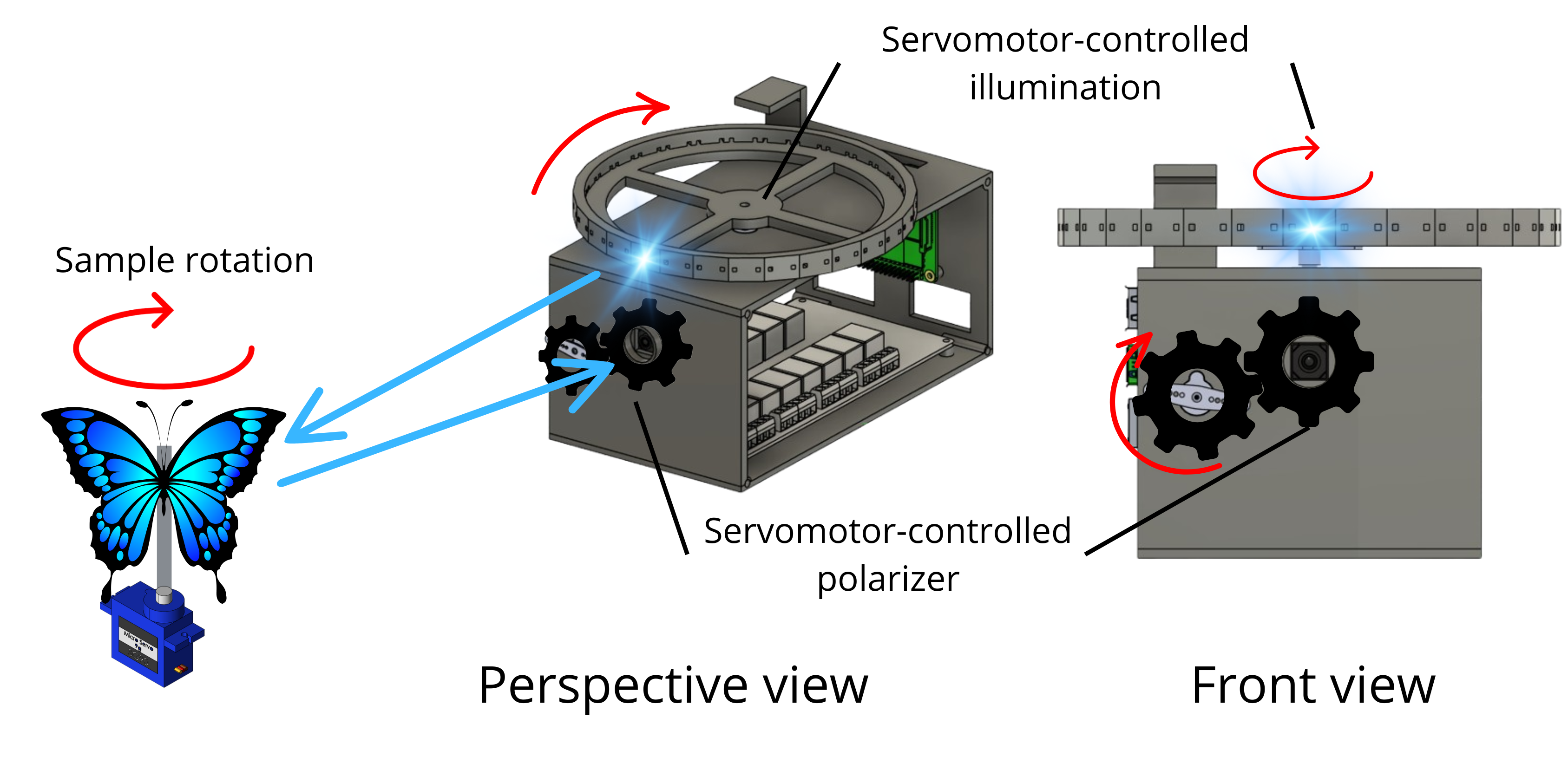}
    \caption{PiMICS for multispectral imaging and polarization scheme. Illumination, polarizer, and sample rotation controlled by a servomotor are shown. The illumination servomotor rotates the 3D printed wheel to position each LED above the camera and in front of the sample for each wavelength. It then takes an image for the polarizer at 0°, rotates it to 90° or in desired increments, and when finished, moves on to the next LED.  The polarizer in front of the LEDs is not shown.}
    \label{epn_scheme}
\end{figure}

The multispectral camera system designed to observe these color behaviors utilizes servomotor-controlled polarization and illumination.  The LEDs are arranged at the radius edge of a 3D printed wheel which is rotated by an SG90 servomotor as shown in fig. \ref{epn_scheme}.  Further, the system incorporates two polarizers. The first fixed polarizer is located in front of the LED light source and the second is in front of the camera module, which is also able to rotate from 0 (co-polarized) to 90 (cross-polarized) degrees in controllable step sizes. 

The LEDs were driven by relay modules (T-Cube LED Driver, Thorlabs, Inc.) set to a 500-mA current source. The logic signals are controlled by the Raspberry Pi, which is connected to an Arduino Nano via I2C communication. The Arduino Nano is responsible for sending the signals to the module.  As a note, this option is considerably more expensive than the transistor option discussed earlier, which works equally well.  

\begin{figure}[ht]
    \centering
    \includegraphics[width=\linewidth]{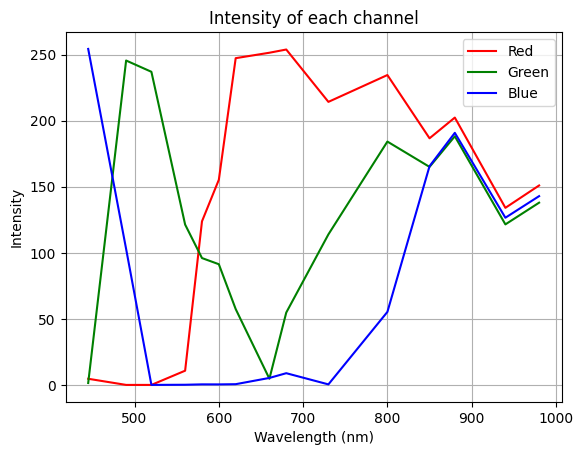}
    \caption{Spectral Response of Raspberry Pi Camera Module 3 - NoIR. Response of the Red (R), Green (G), and Blue (B) channels is shown in the range of LEDs wavelengths.}
    \label{cam_spec_EPN}
\end{figure}

As shown in fig. \ref{cam_spec_EPN} the sensor's sensitivity range spans the visible to about 1000 nm into the near-infrared. This enabled the optimization of light sources and image-channel normalization for accurate interpretation of iridescence intensities and polarization effects.

A PiMICS Console GUI was developed using the Tkinter Python library, with Picamera2 responsible for camera configuration and operation. At present, users are permitted to set the angle of the sampler solely for the purpose of multispectral imaging, whereby two images are captured per wavelength: one co-polarized and one cross-polarized.

The data analysis program processes the background and reference profile, thus facilitating the visualization of intensity profiles for each wavelength, false color images, polar intensity plots, and the generation of a GIF over a loop of wavelengths or rotation angles \cite{HyperPiCodes}.

\subsection{Experimental Results}

\begin{figure}[h]
    \centering
    \includegraphics[width=\linewidth]{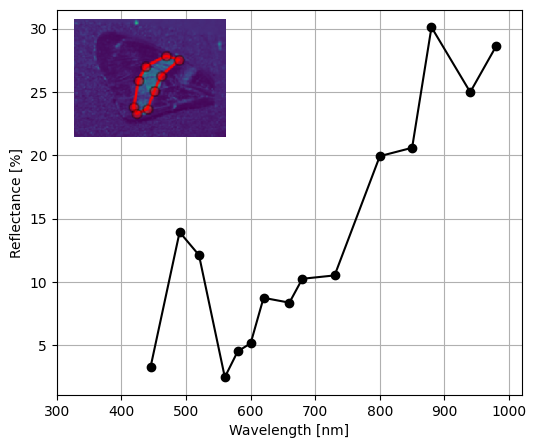}
    \caption{Reflectance spectrum of \textit{Archaeoprepona Demophon} in the co-polarized mode. The region of interest corresponds to the one with observable iridescence. A polygonal mask shown in the upper right was created using the GUI.}
    \label{reflectance_wing}
\end{figure}

Fig. \ref{reflectance_wing} shows multispectral results for \textit{Archaeoprepona Demophon}, a butterfly species native to Mexico and spread throughout northern South America. In addition, the sample was rotated from 0° to 25°, with 5° increments, and two polarization modes were analyzed: co-polarized and cross-polarized.

The spectral reflectance of the region of the butterfly wings where colors characterized by iridescence are observed. The reflectance peak of approximately 15\% occurs around 490 nm, which covers the blue and green region of the visible spectrum. This peak suggests an optimal interaction with these wavelengths, thereby generating an intense color combination. As we move into the near-infrared, the reflectance continues to increase, which is common in biological structures with thin layers or periodic gratings that efficiently scatter light in these regions as a side effect.

Given that the wings of the analyzed species exhibit bright colors within the visible blue region, we investigate whether the intensity of these colors exhibits a behavior dependent on the angle of incidence. 

\begin{figure}[h]
    \centering
    \includegraphics[width=\linewidth]{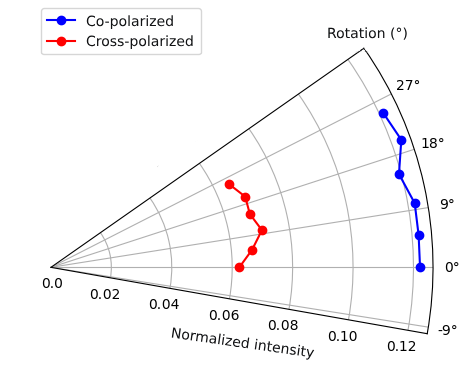}
    \caption{Spectrum of polar intensities of \textit{Archaeoprepona Demophon} for the co-polarized and cross-polarized modes at 445 nm incident light. The intensities are plotted as a function of sample rotation angle, with the data normalized to facilitate comparison.}
    \label{polar_intensities}
\end{figure}

The results presented in fig. \ref{polar_intensities} demonstrate this angular dependence with a favorable reflectance for light of 445 nm at an angle of incidence of 10. This phenomenon creates contrast in the observed colors. According to \cite{LopezAlvarado2022}, the presence of a curvature inside the cavity containing the layers and its anisotropy are responsible for this effect.

\begin{figure}[h]
    \centering
    \includegraphics[width=1\linewidth]{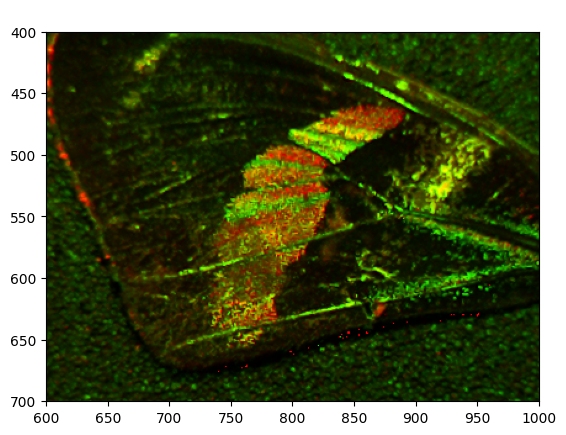}
    \caption{Cross-polarized profiles of the \textit{Archaeoprepona demophon} species wing for 680 nm incident beam and $0^{\circ}$ rotation. The regions where light is detected in cross-polarized mode are highlighted in red.}
    \label{cop-dep}
\end{figure}

Furthermore, the polarization-sensitive interactions that occur and that alter the reflection profile were studied, fig. \ref{cop-dep}. 

The regions indicated in red correspond to those in which cross-polarized light is detected. This indicates that reflected light underwent a linear polarization rotation implying some birefringence in the wing. This polarization-sensitive phenomenon is dependent on both the wavelength and the angle of incidence of the incoming light. As observed, this phenomenon occurs in specific regions with complex nanometric structures, particularly over the area of iridiscence and the borders of the wing. Insects such as butterflies and other species that have polarization-sensitive receptors in their eyes utilize this phenomenon as a means of biological communication.

It should be noted that all results were processed efficiently through the GUI, which allows the exporting of data in tabular format, facilitating analysis for different scientific purposes.

\subsection{Analysis}

As illustrated in Fig. \ref{reflectance_wing}, the maximum reflectance observed within the 490–520 nm range demonstrates the optimized interaction of the wing surface with visible blue and green light. Furthermore, this phenomenon is consistent with the presence of lamellar ridges, which selectively enhance constructive interference for these wavelengths. Therefore, these structures result in structural coloration rather than pigmentation. In addition, these findings are supported by prior research on butterfly wings with analogous morphologies \cite{Wickham2005, wingsiridiscence, wingsIR, LopezAlvarado2022}.

On the other hand, the subsequent decline in reflectance following the 500-nm peak and up to approximately 560 nm is indicative of partial destructive interference, which can be attributed to variations in layer thickness or the presence of additional structures. In contrast, the gradual increase in reflectance beyond 700 nm, extending into the near-infrared (NIR), may serve adaptive purposes. Specifically, the NIR reflectance is a consequence of the structural layers or may serve functional roles, such as thermoregulation by reducing heat absorption in different environmental conditions. In this regard, this is supported by studies indicating that certain butterfly species utilize their wing nanostructures for both optical and thermal purposes \cite{wingsIR}. These results show that nanostructures serve a multifunctional role, providing visual display and environmental adaptability. Further investigations on the behavior of these structures and their anisotropy should be carried out by combining PiMICS with more techniques.

\section{Outreach}
\begin{figure}[ht]
    \includegraphics[width=.46\textwidth]{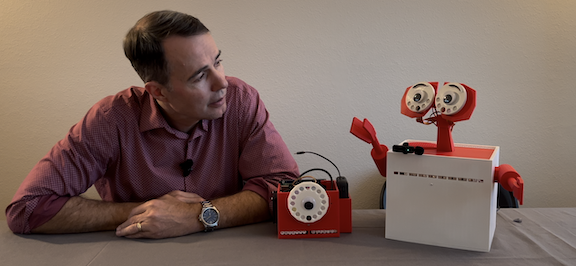}
    \caption{PiMICS 3 doing an interview in which he elaborates on the features of PiMICS Zero. The purpose of building PiMICS 3 is as an outreach tool targeting young scientists. }
    \label{Outreach}    
\end{figure}

Beyond these cameras being used for skills-building and tools for basic research, they are also part of a STEM outreach program.  To make this system interesting as an outreach tool, we made robots (PiMICS 1,2 and 3) with similar features to PiMICS Zero and operated with servomotors.  PiMICS 3 can talk, tell jokes, move, and interact with the audience (see fig. \ref{Outreach} showing PiMICS 3).  It has been a big hit and a great way to teach young people science.  The intent is to use PiMICS 3 to teach about the physics of light, imaging, and spectroscopy. In addition, a YouTube video \cite{PiMICSInterview} and a website \cite{pimics_website} with links to all of the printer files, codes, designs and instructional videos have been developed to better promote and centralize information.     

\section{Conclusion}
In conclusion, we have made a multispectral camera as an educational tool.  The Raspberry Pi-based camera is a low-cost device with broad potential applications.  Students who build the camera will gain a significant number of important skills.  We expect this educational and outreach tool will play a valuable role in developing countries where access to high-end equipment can be difficult and costs prohibitively.  We also believe that programs like this will find wide interest even in developed countries where, all too often, canned experiments dominate the laboratory experience.  

\vfill{\eject}

\bibliography{Main}
% \clearpage
% \appendix
% \onecolumngrid
% \input{supplemental.tex}
\end{document}